%% file: m0416_dmhalos.tex
\documentclass[a4paper,fleqn,usenatbib]{mnras}

\usepackage[T1]{fontenc}
\usepackage{ae,aecompl}

\usepackage{graphicx}	% Including figure files
\usepackage{amsmath}	% Advanced maths commands
\usepackage{amssymb}	% Extra maths symbols
\usepackage{times}
\usepackage{xspace}

\usepackage{color}
\usepackage{hyperref}
\usepackage{supertabular}
\hypersetup{colorlinks, linkcolor=blue}

\def\msun{{\rm M}_{\sun}}

\title[%HFF: Dark Matter \& baryons in strong lensing clusters]{
Shape of galaxy dark matter halos in clusters]{
The shape of galaxy dark matter halos in massive galaxy clusters: Insights from strong gravitational lensing}

\author[Jauzac, Harvey \& Massey\ 2017]{
Mathilde Jauzac,$^{1,2,3,4}$\thanks{E-mail: mathilde.jauzac@durham.ac.uk}
David Harvey,$^{4}$
Richard Massey$^{1,2}$
\newauthor 
\\
$^{1}$Centre for Extragalactic Astronomy, Department of Physics, Durham University, Durham DH1 3LE, U.K.\\
$^{2}$Institute for Computational Cosmology, Durham University, South Road, Durham DH1 3LE, U.K.\\
$^{3}$Astrophysics and Cosmology Research Unit, School of Mathematical Sciences, University of KwaZulu-Natal, Durban 4041, South Africa\\
$^{4}$Laboratoire d'Astrophysique, Ecole Polytechnique F\'ed\'erale de Lausanne (EPFL), Observatoire de Sauverny, CH-1290 Versoix, Switzerland\\
}

\date{Accepted XXX. Received YYY; in original form ZZZ}

\pubyear{2017}

\begin{document}
\label{firstpage}
\pagerange{\pageref{firstpage}--\pageref{lastpage}}
\maketitle

% Abstract of the paper
\begin{abstract}
We assess how much unused strong lensing information is available in the deep \emph{Hubble Space Telescope} imaging and VLT/MUSE spectroscopy of the \emph{Frontier Field} clusters.
As a pilot study, we analyse galaxy cluster MACS\,J0416.1-2403 ($z$$=$$0.397$, $M(R<200\,{\rm kpc})$$=$$1.6$$\times$$10^{14}\msun$), which has 141 multiple images with spectroscopic redshifts.
We find that many additional parameters in a cluster mass model can be constrained, and that adding even small amounts of extra freedom to a model can dramatically improve its figures of merit. We use this information to constrain the distribution of dark matter around cluster member galaxies, simultaneously with the cluster's large-scale mass distribution. 
We find tentative evidence that some galaxies' dark matter has surprisingly similar ellipticity to their stars (unlike in the field, where it is more spherical), but that its orientation is often misaligned. 
When non-coincident dark matter and stellar halos are allowed, the model improves by 35\%.
This technique may provide a new way to investigate the processes and timescales on which dark matter is stripped from galaxies as they fall into a massive cluster.
Our preliminary conclusions will be made more robust by analysing the remaining five \emph{Frontier Field} clusters.
\end{abstract}

% Select between one and six entries from the list of approved keywords.
% Don't make up new ones.

%%%%%%%%%%%%%%%%%%%%%%%%%%%%%%%%%%%%%%%%%%%%%%%%%%

%%%%%%%%%%%%%%%%% BODY OF PAPER %%%%%%%%%%%%%%%%%%
%##################################################%
%##################################################%
\section{Introduction}
\label{sec:intro}

Galaxies are profoundly transformed by infall into a cluster.
As evidenced by the typically smooth distributions of cluster light, infalling subhalos have their gas content efficiently removed to the ICM by ram pressure stripping, even while they pass the virial radius \citep[][]{smith10,wu13}.
In these cluster outskirts, stripping occurs at a rate consistent with simulations. The mass of an $L$* galaxy is reduced by $\sim$$10^{13}\msun$ to $\sim$$10^{12}\msun$ as it falls from a radius of 5\,Mpc to 1\,Mpc \citep[][]{limousin07b,natarajan09}.
However, predictions from simulations disagree when the galaxy continues to the central hundreds of kiloparsecs \citep[e.g.][]{diemand07,penarrubia08,wetzel09}.
The timescale on which dark matter is eventually stripped by tidal forces remains highly uncertain \citep[][]{bahe12}, but the different timescales of infall phases means that the relative ellipticity and alignment between galaxies' stellar and dark matter haloes is likely to change \citep{pereira10}. 
Galaxies in the field are predicted to have dark matter haloes more spherical than their stars, and misaligned semi-major axes \citep{tenneti14,velliscig15}. %Both their shapes and alignment have been measured \citep{gavazzi07,schrabback15,joachimi15}. 
However, the situation is again more complex inside clusters, with different measurement techniques yielding incompatible results \citep[e.g.][]{hao11,sifon15, west17}. 

Galaxy clusters are the largest observable structures in the Universe (e.g.\ Shaye et al.\ 2015).
Comprising of thousands of galaxies embedded within a hot X-ray emitting gas halo and the largest concentrations of mass observed, they have become important testbeds for theories of structure formation \citep{jauzac15a,jauzac16b,jauzac17a,natarajan17} and dark matter \cite[e.g.][]{markevitch04,harvey15}.
The mass of a galaxy cluster can exceed $M>10^{15}M_\odot$ \citep[e.g.][]{umetsu16,jauzac15b,jauzac12,medezinski13,bourdin11}, heavily distorting the curvature of local space-time. As a result, geodesics to objects behind the cluster
become bent, and often split -- resulting in multiple, highly distorted images of background galaxies \citep{bible2,bartelmann10,KN11,hoekstra13}. This effect is known as strong gravitational lensing.
Since the image distortion is directly related to the gradient of the gravitational potential causing it, it is possible to use this information to reconstruct the distribution of 
all matter along the line of sight, including dark matter \cite[e.g.][]{lagattuta17, mahler17, sharon15, richard14}.

This paper attempts, for the first time, to implement a method proposed by \citet[][hereafter H16]{harvey16} to measure infalling galaxies' stellar and dark matter properties using strong gravitational lensing in exceptionally deep, high-resolution observations.
For this, we exploit \emph{Hubble Space Telescope} (HST) Frontier Fields survey \cite[HFF;][]{lotz17} imaging of galaxy cluster MACS\,J0416.1-2403 ($z$$=$$0.397$, $M(R$$<$$200\,{\rm kpc})$$=$$(1.60\pm0.01)$$\times$$10^{14}\,\msun$; \citealt[][hereafter J14]{jauzac14}).
All six HFF clusters have 7-band HST imaging to unprecedented depth.
MACS\,J0416 also has spectroscopy from the \emph{Multi Unit Spectroscopic Explorer} (MUSE) at the \emph{Very Large Telescope} (VLT), which provides spectroscopic redshifts for a large number of strongly-lensed galaxies \citep{caminha17}.
This exceptional information density has led to one of the best-constrained cluster mass models, free from the two usual systematic errors:
(1) the unknown distance between the lensed galaxy and the observer, and (2) the misidentification of counter-images. Using simulations specifically of MACS\,J0416, H16 showed that perturbations of member galaxies' ellipticity, orientation, position and size can shift multiple images by up to 1\arcsec. 
Thus it may be possible to simultaneously constrain the distribution of mass in both the cluster halo and also in individual galaxies.

This paper is organised as follows.
In Sect.~\ref{sec:slmodel} we describe the strong-lensing mass models we use for our analysis.
In Sect.~\ref{sec:results} we present our findings. 
In Sect.~\ref{sec:conclusion}, we discuss their implications. 
Throughout this paper, we assume a $\Lambda$ cold dark matter ($\Lambda$CDM) cosmological model, with $\Omega_m = 0.3$, $\Omega_{\Lambda} = 0.7$, and Hubble constant $H_{0} = 70$\,km\,s$^{-1}$\,Mpc$^{-1}$. 
We note that there currently exists some tension in the measured values of these parameters %and therefore some variation in cosmological parameters may exist
\citep[e.g.][]{riess,planck13_cosmo}.  However, 
while variations in cosmological parameters affect the overall {\em normalisation} of the inferred lens mass, they do not
affect %we do not expect this to affect the relative slope of density profiles, or
the shapes, position angles and positions of halos studied here. %in this study. %, impacting only the overall mass and scaling relation of the cluster. 
Simultaneously constraining cosmological parameters is beyond the scope of this paper.
%In order to explore the actual impact would require an entirely new study and is hence beyond the scope of this paper, and thus we assume a fixed cosmology.
%: $\Omega_m = 0.3$, $\Omega_{\Lambda} = 0.7$, and Hubble constant $H_{0} = 70$\,km\,s$^{-1}$\,Mpc$^{-1}$.}

%##################################################%
%##################################################%
\section{Strong Lensing Mass Models}
\label{sec:slmodel}

%%%%%%%%%%%%%%%%%%%%%%
\subsection{Figures of merit}
This analysis compares various models of the mass distribution in MACS\,J0416. 
To qualitatively evaluate the accuracy of each model we adopt several statistical figures of merit.

Following a Markov Chain Monte Carlo (MCMC) search of model parameter space, we calculate the Bayesian evidence, $E$, and the Bayesian likelihood, $\mathcal{L}$, of the best-fit model.
For these statistics, higher values mean better models.
However, note that none of these figures of merit can compare models with different inputs (such as multiply-imaged galaxies or cluster member catalogues). 
For the maximum likelihood model, we also compute the root-mean-square discrepancy between the predicted and observed positions of lensed galaxies in the image plane, $\langle\mathrm{rms}_i\rangle$.
For this statistic, lower values mean better models.

To compare models with different parameters, we also calculate the Bayesian Information Criterion:
\begin{equation}
{\rm BIC}=-2\log(\mathcal{L})+k\log(N)\ ,
\end{equation}
where $N$ is the number of constraints and $k$ is the number of free parameters.
We then calculate the Akaike Information Criterion: 
\begin{equation}
{\rm AIC}= - 2\log(\mathcal{L}) + 2\ k\ ,
\end{equation}
which is a more robust estimate of overfitting.
We finally consider the Akaike Information Criterion corrected
\begin{equation}
{\rm AICc}={\rm AIC} + \frac{2\ k\ (k+1)}{(N - k - 1)}\ ,
\end{equation}
which corrects the AIC for a finite number of free parameters.
For these figures of merit, lower values should also be preferred.
All three include a penalty term for models with too many free parameters that overfit noise rather than capture additional information.
This penalty term is larger with BIC and AICc than with AIC.
Note that these figures of merit were developed for fits to models with linear parameters. Strong gravitational lensing is highly non-linear, so we report these values but interpret them with caution.

%%%%%%%%%%%%%%%%%%%%%%
\subsection{Fiducial model}

As a fiducial model of the mass distribution in MACS\,J0416, we adopt the CATS model $v4$ from the \emph{Frontier Fields} MAST archive\footnote{\url{https://archive.stsci.edu/pub/hlsp/frontier/macs0416/models/cats/} This is similar to the model in \cite{jauzac14}, but has been re-optimised after redshifts for more multiple images were confirmed by new VLT/MUSE spectroscopy \citep[][]{caminha17}.}. 
The HST observations of MACS\,J0416 were taken under Proposal ID 13396 (PI: Lotz).
This was created using {\sc lenstool} \citep{jullo07} to best reproduce the observed position of 141 multiple images from 51 strongly-lensed galaxies.
At least one of the multiple images in each system has a spectroscopic redshift. Counter-images of each system are either confirmed by a spectroscopic redshift or identified using their geometry, colour and morphology \citep{limousin07b,richard10}. 

\begin{figure*}
\includegraphics[width=\textwidth]{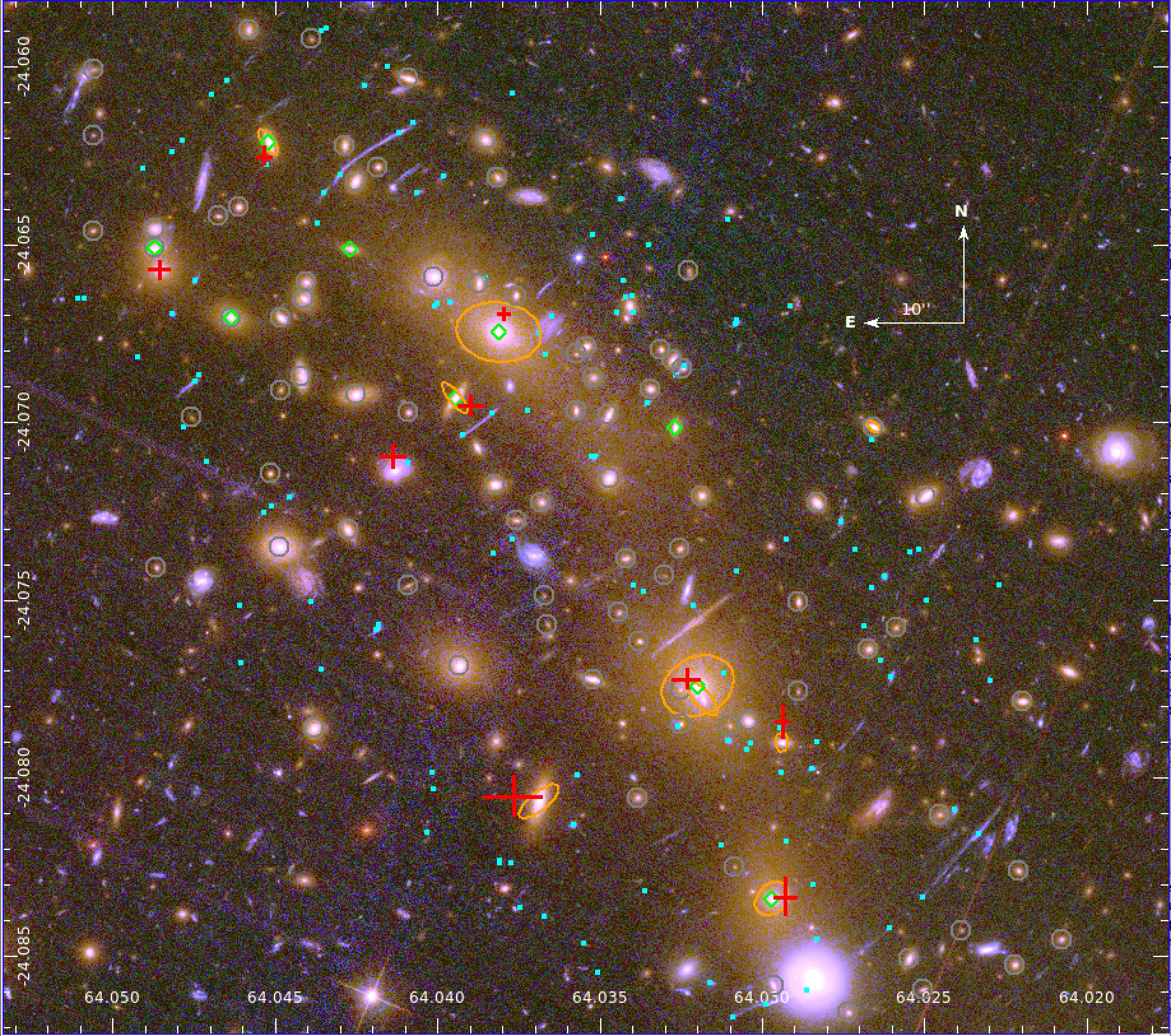}
\caption{Composite HST/ACS colour image of MACS\,J0416 (axes show right ascension and declination in degrees). 
Grey circles show the cluster galaxies included in our mass models, assuming the \citet{FJ76} relation between luminosity and mass. 
Green diamonds highlight the 9 galaxies individually optimized for the `Ellipticity' model.
Orange ellipses highlight the 9 galaxies individually optimized for the `Position Angle' model: the ellipse is oriented following the best-fit value of $\theta$. The red crosses show the 9 galaxies for which the dark matter halo position ($x$,$y$) is constrained: they are centered on the best-fit position of the dark matter halo, the size of the bars correspond to the 95\% error on that position. 
Cyan squares show the positions of strongly-lensed images of background galaxies that we used to optimise our models.}
\label{dmoffset}
\end{figure*}

The cluster mass distribution is built from 3 cluster-scale halos. We include 96 member galaxies with mass-to-light ratios fixed by the \cite{FJ76} relation (grey circles in Fig.~\ref{dmoffset}). Two more
galaxies close to multiple images are modelled individually, i.e. not assuming the \cite{FJ76} relation, as they are showing clear signs of galaxy-galaxy lensing.
Every component is modelled as a pseudo-isothermal elliptical mass distribution \citep[PIEMD;][]{eliasdottir09}, characterized by a spatial position ($x$,$y$), ellipticity $\epsilon$, orientation $\theta$, core radius $r_\mathrm{core}$, scale radius $r_\mathrm{s}$, and velocity dispersion $\sigma$. 

In previously published mass models, galaxies are represented by a single halo containing all of their mass. Parameters describing the shape and size of each halo are fixed from the distribution of light in the F814W-band; the velocity dispersion is fixed from the F814W-band flux, such that a galaxy's total mass follows the \cite{FJ76} relation.
For the first time, we separate galaxies into components of dark and stellar matter.
We preserve total mass, and fix stellar mass to values measured by the {\sc astrodeep} survey, using Spectral Energy Density (SED) fitting to multiband photometry \citep{merlin16,castellano16}.
This photometry includes 7 optical and near infrared bands from HST/ACS and VLT/Hawk-I, plus Spitzer/IRAC 3.6 and 4.5 $\mu$m bands. The SED fitting measures physical properties using six different methods to minimise residual systematics.
%We preserve total mass, and fix stellar mass from {\sc astrodeep} fits to multi-wavelength photometry \citep{merlin16,castellano16}.
%\textbf{They derive photometric redshifts and other physical parameters using 7 optical and near infrared bands from HST and publicly available Hawk-I on the VLT and IRAC 3.6 and 4.5 $\mu$m bands. They use six different methods to measure the physical properties to ensure that no residual systematics exist. For more information please see \cite{merlin16} and \cite{castellano16}.}
Initially, we assume the two components are coincident, so the model has no extra free parameters. Figures of merit for this model are listed in Table~\ref{tab:results}. 

%%%%%%%%%%%%%%%%%%%%%%
\subsection{Models with additional free parameters} 

We shall now model the distribution of dark matter in some galaxies independently of their distribution of stars.
With a parameter space spanning up to 7 parameters for each of 96 galaxies, optimisation could rapidly become impossible.
To make the calculation tractable, we prioritise galaxies to model individually.

We first identify cluster member galaxies whose distribution of dark matter might be constrainable without prior bias.
To be conservative in this proof-of-concept analysis, we exclude from consideration galaxies with ellipticity $\epsilon<0.2$ or stellar-to-halo-mass ratio $\mathrm{SHMR}>0.3$.
The first cut is because of the way {\sc lenstool} parameterises ellipticity as $(\epsilon,\theta)$, with a strict bound $\epsilon>0$.  
For example, after randomly perturbing the positions of multiple images, galaxies that are nearly circular appear to gain a best-fit ellipticity that is biased high by this bound.
The second cut is because scatter in the \cite{FJ76} relation leads to unphysical values of our simple SHMR estimates, which prevent us from splitting a galaxy's total mass between stellar and dark matter components in a consistent way.
These cuts lead to a catalogue of 29 galaxies, spread fairly uniformly throughout the cluster.
Both limitations could be avoided in principle, and a few additional galaxies could be considered in future work. 

\input{table1}

\begin{figure}
\centering
\includegraphics[width=0.25\textwidth]{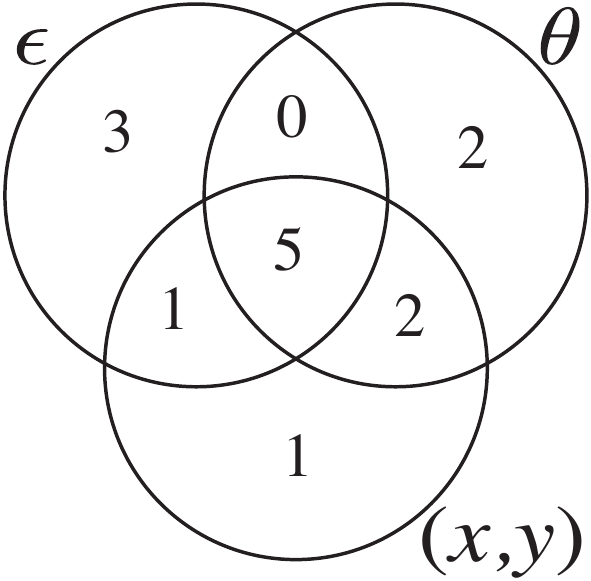}
\caption{Venn diagram showing the overlap between the three sets of 9 galaxies being considered for ellipticity, orientation and spatial position optimization.
}
\label{venndiag}
\end{figure}

To identify galaxies whose distribution of dark matter will be constrained with statistical significance, H16 suggests finding all those near multiple images -- but notes that some configurations of strong lensing are more constraining than others.
To avoid missing any constrainable galaxies, we implement a two-step process. 
First, for all 29 eligible galaxies, we use {\sc lenstool} to vary the ellipticity $\epsilon$, orientation $\theta$ and position $(x,y)$ of dark matter within broad priors. 
The parameter space is highly dimensional, so the posterior probability density function (PDF) remains noisy after any reasonable amount of computing time.
However, we can predict which parameters will be constrainable in a second optimisation, by fitting Gaussians to the noisy PDF.
We identify nine galaxies whose 1D PDF has width $\sigma_\epsilon<0.3$, nine galaxies with $\sigma_\theta<60^\circ$, and nine galaxies with $\sigma_x<2\arcsec$ and $\sigma_y<2\arcsec$.
They are not the same nine galaxies in each case, because different lensing configurations constrain different properties of a local mass distribution (it is a coincidence that each requirement leads to nine galaxies).
The galaxies in each set are shown by the different symbols in Figure~\ref{dmoffset} (green diamonds for ellipticity, orange ellipses for position angle and red crosses for position). Their coordinates and the overlap between sets are listed in Table~\ref{tab:gal}. The overlap is also illustrated in Figure~\ref{venndiag}.

Finally, we re-initialise the MCMC for three last optimisations.
The cluster-scale halos are being optimized with the two galaxies responsible for galaxy-galaxy lensing, galaxy-scale halos are optimized following the \cite{FJ76} relation, however for nine galaxies simultaneously, we optimise either the dark matter
ellipticity $\epsilon$, orientation $\theta$, or spatial position ($x$,$y$). 

%##################################################%
%##################################################%
\section{Results}
\label{sec:results}
We shall now discuss the results of each re-optimisation, and compare each best-fit model with the fiducial one. Figures of merit for these models are listed in Table~\ref{tab:results}.

\input{table2}
%%%%%%%%%%%%%%%%%%%%%%
\subsection{Varying the ellipticity of galaxies' dark matter}

Allowing nine galaxies' dark matter halos to have a different axis ratio than its stars improves the best-fit 
$\langle\mathrm{rms}_i\rangle$ by $\sim13\%$ compared to the fiducial model. 
The 2\% decrease in BIC and 6\% decrease in AICc suggest that this improvement is not simply due to overfitting noise, but reflects real inadequacy in the fiducial model.

\begin{figure}
\includegraphics[width=0.47\textwidth]{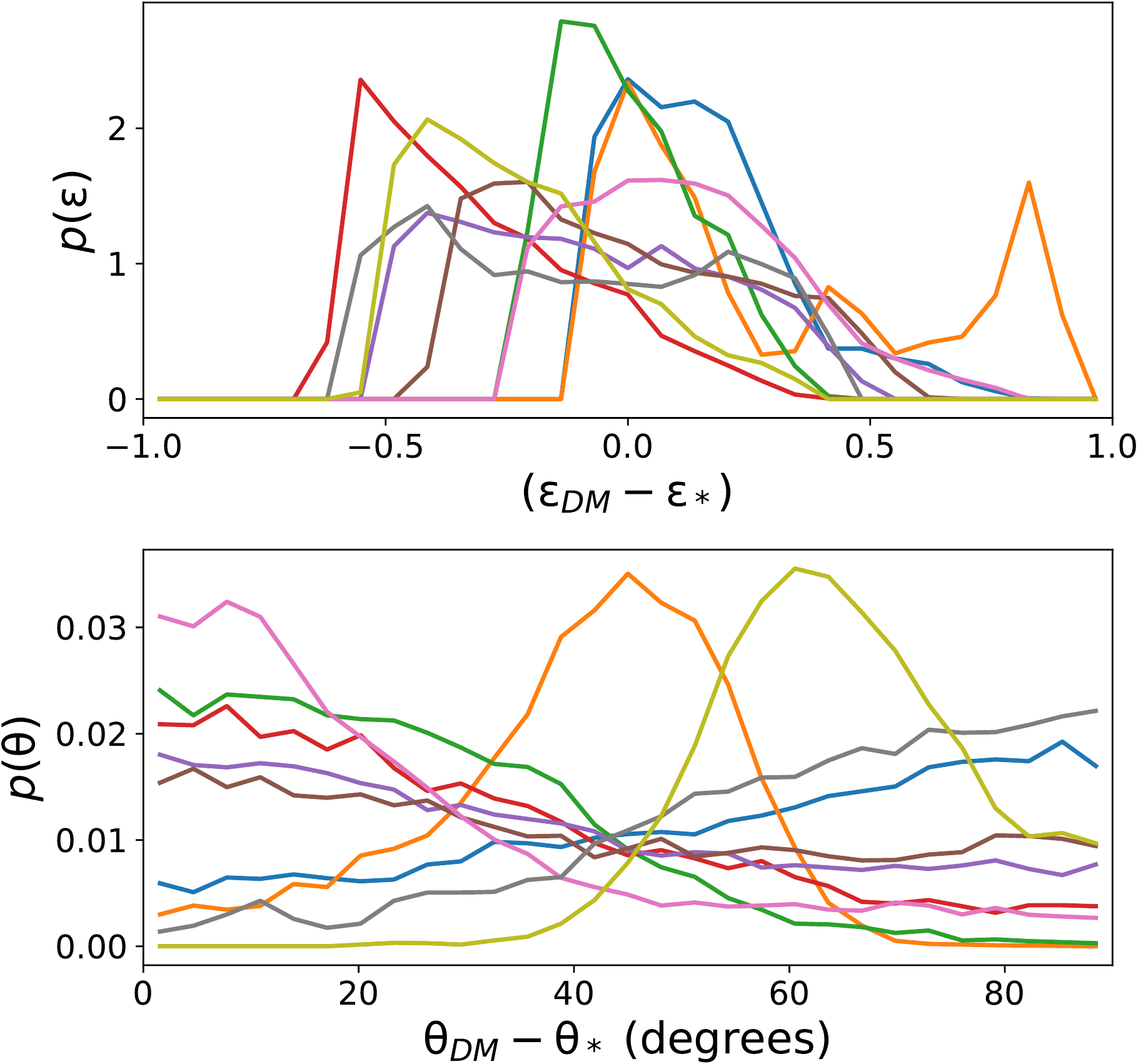}
\caption{\emph{Top Panel:} Probability distribution of the ellipticity as a function of the difference between the ellipticity of the stellar halo and the ellipticity of the dark matter halo for the 9 galaxies that have constrained ellipticities from the model. The black line corresponds to the mean over all galaxies.
\emph{Bottom Panel:} Probability distribution of the acute angle $\theta$ between the major axes of a galaxy's stars and its dark matter.
}
\label{angposell}
\end{figure}

We find that about half the galaxies have dark matter that is more spherical than the stars, as expected (Figure~\ref{angposell}).
However, the other half of the galaxies have dark matter with ellipticity similar to that of their stellar haloes, in contradiction with numerical simulations which predict dark matter halos to be more spherical than the stellar ones \citep[e.g.][]{velliscig15,tenneti14}.
Note that the extreme end of one or two galaxies' PDFs in Fig.~\ref{angposell} may be truncated by priors, but this effect appears robust to excluding those.
Multiplying the PDFs together, to represent the net galaxy population, yields a mean value $\langle\epsilon_\mathrm{DM}-\epsilon_\star\rangle=0.01\pm0.05$. 

A subtlety of ellipticity measurement is that parametric fits (as obtained from {\sc lenstool}) give $\sim10\%$ higher absolute values for the same distribution than moment-based measurements (as obtained for {\sc astrodeep} using \textsc{sextractor}; \citealt{BA96}).
For the fairest comparison, this analysis therefore reports stellar ellipticities instead measured using the \textsc{ciao 4.9} fitting tool \textsc{sherpa} \citep{freeman01}. This does not change our qualitative conclusion.

%%%%%%%%%%%%%%%%%%%%%%
\subsection{Varying the orientation of galaxies' dark matter}

Allowing nine galaxies' dark matter halos to be rotated with respect to the stars improves 
$\langle\mathrm{rms}_i\rangle$ by $\sim9\%$ compared to the fiducial model. 
The decrease in BIC and AICc is similar to the previous test.

We find that the dark matter in about half (5/9) of the galaxies is aligned with the stars. 
However, dark matter in two of the galaxies is significantly misaligned with ($\sim45^\circ$ from) the stars, and in two of the galaxies it is consistent with being maximally misaligned by $90^\circ$. 
The net galaxy population has a mean misalignment $\langle\theta_\mathrm{DM}-\theta_\star\rangle=48\pm8$. 

%%%%%%%%%%%%%%%%%%%%%%
\subsection{Varying the position of galaxies' dark matter}

Allowing nine galaxies' dark matter halos to be spatially offset from the stars adds twice as many free parameters as the other tests, but improves $\langle\mathrm{rms}_i\rangle$ by $\sim35\%$ compared to the fiducial model. 
BIC and AICc decrease by 19\% and 28\%, again suggesting that a more complex model is not simply fitting noise.
However, individual galaxies' PDFs remain broad, and it is difficult to quantify the error in the position of a dark matter halo due to the complex configuration of multiple images \citep{harvey17b}.
There is no net preferred direction along which to average spatial offsets from the entire sample.

%##################################################%
%##################################################%
\section{Discussion \& Future Work} 
\label{sec:conclusion}

Parametric models are widely used to fit the mass distribution in galaxy clusters.
They are the most successful at reproducing observed positions of multiple images in simulated clusters \citep{meneghetti17}, and ostensibly reach a statistical precision of 1\% .
These models are exploited to study physical processes in the cluster itself \citep[e.g.][]{jauzac15a,ogrean15,annunziatella17} or to use its gravitational lensing as a natural telescope to observe the high redshift Universe \citep[e.g.][]{atek15b,ishigaki15,kawamata16}.
We have demonstrated that the latest high resolution, deep images of galaxy clusters contain a lot more information than is accounted for by parametric models.
Commonly-used figures of merit such as $\langle\mathrm{rms}_i\rangle$ can be improved dramatically by the introduction of relatively few extra free parameters.

If the parameters are chosen well, the additional information content could be used to model the distribution of mass in cluster member galaxies.
One tentative, but potentially interesting result from our analysis is that some cluster galaxies appear to have an equally high dark matter halos axis ratio as their stellar companions.
In field galaxies, the distribution of dark matter is more spherical than the stars.
This discrepancy could potentially help to constrain the mechanisms and timescales for the stripping of dark matter from galaxies during infall into clusters. On the other hand, we find that the orientation of galaxy's dark matter halos is often misaligned, which could also provide an alternate probe to the physics of harassment during galaxy infall.

Alternatively, the additional information could indicate (and quantify) deficiencies in the models used to fit galaxy clusters.
We also find that our model of MACS\,J0416 dramatically improves by 35\% when we let the position of the dark matter to separate from that of the stars. This improvement however has no common trajectory or obvious physical origin so we therefore attribute this to the lack of complexity for the cluster-scale halos imposed by the parametric approach.
Parametric models can span only a limited range of mass distributions, and cannot capture the full complexity seen in simulations around the most massive structures in the Universe.
While their statistical precision may approach 1\%, their accuracy may not be as good.

We are optimistically inclined to believe that the information is truly connected to galaxy properties, because the improvements in fit come from all over the cluster rather than one (perhaps poorly modelled) region.
One way to distinguish between these scenarios will be to repeat our test, but starting from free-form mass models \cite[e.g.][]{diego14} that already have sufficient flexibility to capture more substructures.
It would then be necessary to add individual halos for each galaxy that are as orthogonal as possible to existing free parameters.

It will also be necessary to expand our analysis to more galaxy clusters.
To complete a proof-of-concept, this paper analysed a single \emph{Hubble Frontier Field} cluster, MACS\,J0416.
The limited statistical significance of our results does not yet support robust conclusions.
However, the surprisingly large improvement in figures of merit such as $\langle\mathrm{rms}_i\rangle$ clearly demonstrates the available information content that is being missed by current analyses, and justifies an expansion of the study.
At best, this information will provide a useful new tool to investigate the properties of galaxies' dark matter, as they are transformed by their infall into massive clusters.
At worst, better statistics will quantify systematic effects in current studies of the high redshift, gravitationally lensed Universe.

%##################################################%
%##################################################%
\section*{Acknowledgements}
The authors thank the anonymous referee for its useful comments that helped improved this paper.
The authors thank J. Richard, M. Schaller and A. Robertson for fruitful discussions.
This work was supported by the Science and Technology Facilities Council [grant numbers ST/L00075X/1 and ST/P000541/1], and a 2016 Durham University International Engagement Grant.
This work used the DiRAC Data Centric system at Durham University, which is operated by the Institute for Computational Cosmology on behalf of the STFC DiRAC HPC Facility (\url{www.dirac.ac.uk}). This equipment was funded by BIS National E-infrastructure capital grant ST/K00042X/1, STFC capital grant ST/H008519/1, and STFC DiRAC Operations grant ST/K003267/1 and Durham University. DiRAC is part of the National E-Infrastructure. This research was supported by the Swiss National Science Foundation (SNSF).
MJ acknowledges the M\'esocentre d'Aix-Marseille Universit\'e (project number: 16b030).  
MJ acknowledges support from the ERC advanced grant LIDA. 
DH acknowledges support by the Merac foundation.
RM is supported by the Royal Society.
Based on observations obtained with the NASA/ESA Hubble Space Telescope, retrieved from the Mikulski Archive for Space Telescopes (MAST) at the Space Telescope Science Institute (STScI). STScI is operated by the Association of Universities for Research in Astronomy, Inc. under NASA contract NAS 5-26555. 

%%%%%%%%%%%%%%%%%%%%%%%%%%%%%%%%%%%%%%%%%%%%%%%%%%

%%%%%%%%%%%%%%%%%%%% REFERENCES %%%%%%%%%%%%%%%%%%

% The best way to enter references is to use BibTeX:

\bibliographystyle{mnras}
\bibliography{reference} % if your bibtex file is called example.bib

%%%%%%%%%%%%%%%%%%%%%%%%%%%%%%%%%%%%%%%%%%%%%%%%%%

%%%%%%%%%%%%%%%%% APPENDICES %%%%%%%%%%%%%%%%%%%%%

%\appendix
%
%\section{Some extra material}
%
%If you want to present additional material which would interrupt the flow of the main paper,
%it can be placed in an Appendix which appears after the list of references.

%%%%%%%%%%%%%%%%%%%%%%%%%%%%%%%%%%%%%%%%%%%%%%%%%%

% Don't change these lines
\bsp	% typesetting comment
\label{lastpage}
\end{document}

%% file: table1.tex
\begin{table}
\caption{Coordinates of the individually optimised galaxies. We also highlight which one is being optimised by which model(s).
}
%    \begin{minipage}{0.72\textwidth}
%\begin{center}
%\begin{flushright}
\begin{tabular}{ccccc}
\hline
\hline
R.A. & Dec. & Ellipticity & Position & Position\\
($deg$) & ($deg$) & & Angle & \\
%\\
\hline
\hline
%247 & 
64.02969360 & -24.08340836 & X & X & X \\
%285 & 
64.03686523 & -24.08066559 & & X & X \\
%310 & 
64.03196716 & -24.07742882 & X & X & X \\
%357 & 
64.03182220 & -24.07779312 & & X & \\
%361 & 
64.02938080 & -24.07900810 & & X & X \\
%468 & 
64.03810119 & -24.06748390 & X & X & X \\
%478 & 
64.04131667 & -24.07160778 & & & X \\
%504 & 
64.03945160 & -24.06933212 & X & X & X \\
%508 & 
64.03267670 & -24.07014847 & X & & \\
%520 & 
64.02655792 & -24.07013321 & & X & \\
%557 & 
64.04636383 & -24.06705856 & X & & \\
%601 & 
64.04871368 & -24.06511497 & X & & X \\
%616 & 
64.04269409 & -24.06512833 & X & & \\
%645 & 
64.04518890 & -24.06213570 & X & X & X \\

\hline
\hline
\end{tabular}
\label{tab:gal}
%\end{flushright}
%%\end{center}
%    \end{minipage}%
%    \begin{minipage}{0.18\textwidth}
%        \centering
%        \includegraphics[width=3cm]{VennDiagram_v2.pdf}
%    \end{minipage}
\end{table}

%% file: table2.tex
\begin{table*}
\caption{Figures of merit for each model considered in this paper. Columns show the number of degrees of freedom ($d.o.f.$), the number of free parameters ($k$), the total number of parameters ($N$), Bayesian evidence ($\log{E}$) and likelihood ($\log{\mathcal{L}}$), the rms deviation of predicted multiple-image positions from their observed positions in the image plane $\langle\mathrm{rms}_i\rangle$, the Bayesian Information Criterion, the Akaike Information Criterion, and the corrected AIC.
We also quote the improvement on several parameters compare to the fiducial model: on the $\langle\mathrm{rms}_i\rangle$, $\delta_{\mathrm{rms}_i}$, the BIC, $\delta_{\rm BIC}$, and the AICc, $\delta_{\rm AICc}$. A value of $\delta_{\rm BIC}$ and $\delta_{\rm AICc}$ greater than 10 reflects a significant improvement of the model compare to the fiducial one.
%The Venn diagram shows the number of galaxies whose parameters are individually constrained in each of the three models.
}
%    \begin{minipage}{0.72\textwidth}
%\begin{center}
%\begin{flushright}
\begin{tabular}{ccccccccccccc}
\hline
\hline
Model & $d.o.f.$ & $k$ & $N$ & $\log{E}$ & $\log{\mathcal{L}}$ & $\langle\mathrm{rms}_i\rangle$ & BIC & AIC & AICc & $\delta_{\mathrm{rms}_i}$ & $\delta_{\rm BIC}$ & $\delta_{\rm AICc}$\\
%\\
\hline
\hline
Fiducial & 154 & 26 & 180 & -493.07 & -245.91 & 0.80\arcsec & 627 & 543 & 553 & -- & -- & -- \\
\hline
%Ellipticity & 125 & 55 & 180 & -267.11 & -202.80 & 0.70\arcsec & 691 & 515 & 565 \\
%Position Angle & 125 & 55 & 180 & -256.33 & -213.30 & 0.73\arcsec & 712 & 536 & 586 \\
%Position ($x$,\,$y$) & 96 & 84 & 180 & -194.66 & -136.15 & 0.51\arcsec & 708 & 440 & 590 \\
%\hline
Ellipticity & 145 & 35 & 180 & -280.24 & -216.81 & 0.74\arcsec & 615 & 503 & 521 & 8\% & 12 & 32 \\
Position Angle & 145 & 35 & 180 & -271.08 & -217.76 & 0.74\arcsec & 617 & 506 & 523 & 8\% & 10 & 30 \\
Position ($x$,\,$y$) & 136 & 44 & 180 & -209.49 & -140.29 & 0.52\arcsec & 509 & 369 & 398 & 35\% & 118 & 155 \\
\hline
\hline
\end{tabular}
\label{tab:results}
%\end{flushright}
%%\end{center}
%    \end{minipage}%
%    \begin{minipage}{0.18\textwidth}
%        \centering
%        \includegraphics[width=3cm]{VennDiagram_v2.pdf}
%    \end{minipage}
\end{table*}